\documentclass[preprint,showpacs,preprintnumbers,amsmath,amssymb]{revtex4-1}

\usepackage{graphicx}
\usepackage{dcolumn}
\usepackage{bm}

\begin{document}

\title{Diffraction phenomenology with massive gluons: \\ some recent developments}

\author{E.G.S. Luna}
\affiliation{
Instituto de F\'{\i}sica e Matem\'atica, Universidade Federal de Pelotas, 96010-900, Pelotas, RS, Brazil}


\begin{abstract}
In this talk we introduce the main features of a QCD-based model in which the coupling $\alpha_{s}$ is constrained by an infrared mass scale.
We show recent applications of this model to hadron-hadron collisions, gap survival probability calculations, and soft gluon resummation
techniques. These results indicate a smooth transition from nonperturbative to perturbative behaviour of the QCD.
\end{abstract}

\maketitle

\section{Introduction}

At high energies the soft and the semihard components of the scattering amplitude are
closely related \cite{gribov,ryskin01}, and it becomes important to distinguish between semihard gluons, which participate in hard
parton-parton scattering, and soft gluons, emitted in any given parton-parton QCD radiation process. A class of models based on QCD
incorporate soft
and semihard processes in the treatment of high-energy hadronic interactions using a formulation compatible with analyticity and
unitarity constraints \cite{godbole,godbole2,durand1,durand2,luna01}. In this talk we present a QCD-based model \cite{luna01} in which the
coupling $\alpha_{s}$ is constrained by the value
of the so called ``dynamical gluon mass'', whose existence is strongly supported by recent QCD lattice simulations. This frozen coupling,
obtained by means of the pinch technique in order to derive
a gauge invariant Schwinger-Dyson equation for the gluon propagator \cite{cornwall1,cornwall2}, has been adopted in many phenomenological
studies \cite{luna01,halzen,halzen2,natale01,natale02,luna02,luna03,luna04,luna05}. More
specifically, we discuss
some recent applications of the model to hadron-hadron collisions, gap survival probability calculations, and soft gluon resummation
techniques. These results indicate a smooth transition from nonperturbative to perturbative behaviour of the QCD.

\section{Hadron-hadron collisions}

In the QCD-based (or ``mini-jet") models the increase of the total cross sections is associated
with semihard scatterings of partons in the hadrons \cite{godbole,godbole2,durand1,durand2,luna01}. The energy dependence of the cross sections is
driven especially by gluon-gluon scattering processes, where the behaviour of the gluon
distribution function at small $x$ exhibits the power law $g(x,Q^{2}) \sim x^{-J}$. Since
gluon-gluon subprocesses are potentially divergent at small transferred
momenta, it becomes important to regulate this behaviour by introducing a mass
scale which separates the perturbative from the non-perturbative QCD region. In the so called ``DGM'' model (Dynamical Gluon Mass model) \cite{luna01},
the dynamical gluon mass, as well as the infrared finite coupling constant associated to it \cite{ans}, are the
natural regulators for the cross sections calculations. In this eikonal model, the total cross section, the ratio
$\rho$ of the real to the imaginary part of the forward scattering amplitude, and the differential elastic scattering cross
section are given by
\begin{eqnarray}
\sigma_{tot}(s) = 4\pi \int_{_{0}}^{^{\infty}} \!\! b\, db\, [1-e^{-\chi_{_{I}}(b,s)}\cos \chi_{_{R}}(b,s)],
\label{degt1}
\end{eqnarray}
\begin{eqnarray}
\rho (s) = \frac{\textnormal{Re} \{ i \int b\, db\, [1-e^{i\chi (b,s)}]  \}}{\textnormal{Im} \{ i \int b\,
db\, [1-e^{i\chi (b,s)}]  \}},
\label{degthyj1}
\end{eqnarray}
and
\begin{eqnarray}
\frac{d\sigma_{el}}{dt}(s,t)=\frac{1}{2\pi}\, \left| \int b\, db\, [1-e^{i\chi (b,s)}]\, J_{0}(qb) \right|^2 ,
\label{degthyj3}
\end{eqnarray}
respectively, where $s$ is the square of the total CM energy, and
$\chi(b,s)=\chi_{_{R}}(b,s)+i\chi_{_{I}}(b,s)$ is the (complex) eikonal function, which is written as a combination of an even and odd
eikonal terms related by crossing symmetry. In terms of the proton-proton ($pp$) and
antiproton-proton ($\bar{p}p$) scatterings, this combination reads
$\chi_{pp}^{\bar{p}p}(b,s) = \chi^{+} (b,s) \pm \chi^{-} (b,s)$. 
The even eikonal is written as the sum of gluon-gluon, quark-gluon, and quark-quark contributions:
\begin{eqnarray}
\chi^{+}(b,s) &=& \chi_{qq} (b,s) +\chi_{qg} (b,s) + \chi_{gg} (b,s) \nonumber \\
&=& i[\sigma_{qq}(s) W(b;\mu_{qq}) + \sigma_{qg}(s) W(b;\mu_{qg})+ \sigma_{gg}(s) W(b;\mu_{gg})] ,
\label{final4}
\end{eqnarray}
where $W(b;\mu)$ is the overlap
function at impact parameter space and $\sigma_{ij}(s)$ are the elementary subprocess cross sections of colliding quarks and
gluons ($i,j=q,g$). The odd eikonal $\chi^{-}(b,s)$, that
accounts for the difference between $pp$ and $\bar{p}p$ channels, is parametrized as
\begin{eqnarray}
\chi^{-} (b,s) = C^{-}\, \Sigma \, \frac{m_{g}}{\sqrt{s}} \, e^{i\pi /4}\, 
W(b;\mu^{-}),
\label{oddeik}
\end{eqnarray}
where $m_{g}$ is the dynamical gluon mass and the parameters $C^{-}$ and $\mu^{-}$ are constants to be
fitted. The factor $\Sigma$ is defined as $\Sigma = \frac{9\pi \bar{\alpha}_{s}^{2}(0)}{m_{g}^{2}}$, with the dynamical coupling constant
$\bar{\alpha}_{s}$ set at its frozen infrared value. The gluon-gluon eikonal
contribution, that dominates at high energy and determines the asymptotic behaviour of the total cross sections, is written as
$\chi_{gg}(b,s)\equiv \sigma_{gg}^{D\!PT}(s)W(b; \mu_{gg})$, where
\begin{eqnarray}
\sigma_{gg}^{D\!PT}(s) = C^{\prime} \int_{4m_{g}^{2}/s}^{1} d\tau \,F_{gg}(\tau)\,
\hat{\sigma}^{D\!PT}_{gg} (\hat{s}) .
\label{sloh1}
\end{eqnarray}
Here $F_{gg}(\tau)$ is the convoluted structure function for pair $gg$, $\hat{\sigma}^{D\!PT}_{gg}(\hat{s})$ is
the subprocess cross section and $C^{\prime}$ is a fitting parameter. In the
above expression it is introduced the energy threshold $\hat{s}\geq 4m_{g}^{2}$ for the final state gluons,
assuming that these are screened gluons \cite{cornwall1,cornwall2}. In the expression (\ref{sloh1}) the gluon-gluon subprocess cross
section $\hat{\sigma}^{D\!PT}_{gg}(\hat{s})$ is calculated
using a procedure dictated by the dynamical perturbation theory (DPT) \cite{pagels}. It means that the effects of the
dynamical gluon mass in the propagators and vertices
are retained, and the sum of polarizations is performed for massless (free-field) gluons. As a result, since the dynamical
masses go to zero at large momenta, the elementary cross sections of perturbative QCD in the high-energy limit are recovered \cite{luna01}.
The result of the fit to $\sigma_{tot}$ for both $pp$ and $\bar{p}p$ channels is displayed in Figs. 1, together with the experimental data.
\begin{figure}
 \includegraphics[width=0.37\columnwidth]{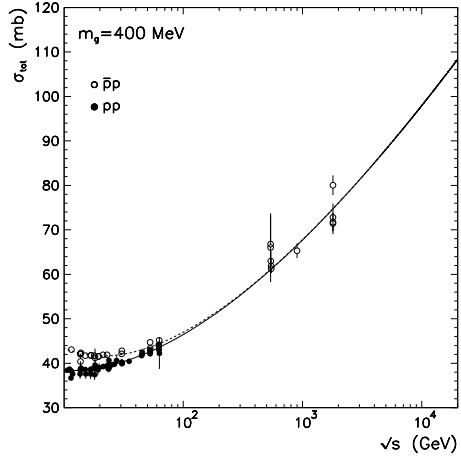}
\caption{Total cross section for $pp$ (solid curve) and $\bar{p}p$ (dashed curve) scattering.}
\end{figure}

\section{Survival probability of large rapidity gaps}

The study of the survival probability $\langle |S|^{2} \rangle$ of large rapidity gaps (LRG) is currently a subject of
intense theoretical and experimental interest. Its importance lies in the fact that systematic analyses of
LRG open the possibility of extracting New Physics from hard diffractive processes. On the theoretical side, its
significance is due to the reliance of the $\langle |S|^{2} \rangle$ calculation on subtle QCD methods. Since rapidity gaps can occur in the case of Higgs
boson production via fusion of electroweak bosons, we have focused on $WW\to H$ fusion processes \cite{luna02}. The inclusive differential
Higgs boson production cross section via $W$ fusion is given by
\begin{eqnarray}
\frac{d\sigma_{prod}}{d^{2}\vec{b}}=\sigma_{WW\to H}\, W(b;\mu_{W} ),
\label{eqnarrat01}
\end{eqnarray}
where $W(b;\mu_{W})$ is the overlap function at impact parameter space of the $W$ bosons. This function represents the effective
density of the overlapping $W$ boson distributions in the colliding hadrons. The cross section for producing the Higgs
boson and having a large rapidity gap is given by
\begin{eqnarray}
\frac{d\sigma_{LRG}}{d^{2}\vec{b}} &=& \sigma_{WW\to H}\, W(b;\mu_{W} )\, P(b,s),
\label{eqnarrat02}
\end{eqnarray}
where $P(b,s)$ is the probability that the two initial hadrons have not undergone a
inelastic scattering at the parton level. In the DGM model this probability is given by
$P(b,s)=e^{-2\chi_{I}(b,s)}$, where the imaginary part $\chi_{I}(b,s)$ of the eikonal function receives contributions of
parton-parton interactions. Therefore, the factor $P(b,s)$
suppresses the contribution to the Higgs boson cross section where the two initial hadrons overlap and there is soft
rescatterings of the spectator partons. Hence in our model we can write down the survival factor $\langle |S|^{2} \rangle$ for Higgs production via $W$ fusion
as
\begin{eqnarray}
\langle |S|^{2} \rangle = \frac{\int d^{2}\vec{b} \, \sigma_{WW\to H}\, W(b;\mu_{W} )\,
e^{-2\chi_{I}(b,s)}}{\int d^{2}\vec{b} \,
\sigma_{WW\to H}\, W(b;\mu_{W} )} = \int d^{2}\vec{b} \, W(b;\mu_{W} )\, e^{-2\chi_{I}(b,s)} ,
\label{eqnarrat03}
\end{eqnarray}
where we have used the normalization condition $\int d^{2}\vec{b} \, W(b;\mu_{W} ) = 1$. The sensitivity of the survival probability
$\langle |S|^{2} \rangle$ (for $pp$ collisions) to the gluon dynamical mass is
shown in Figure 1 for some CM energies. The $\langle |S|^{2} \rangle$ value decreases with the increase of the energy of the incoming hadrons, in line
with the available experimental data for LRG.
\begin{figure}
\includegraphics[width=0.37\columnwidth]{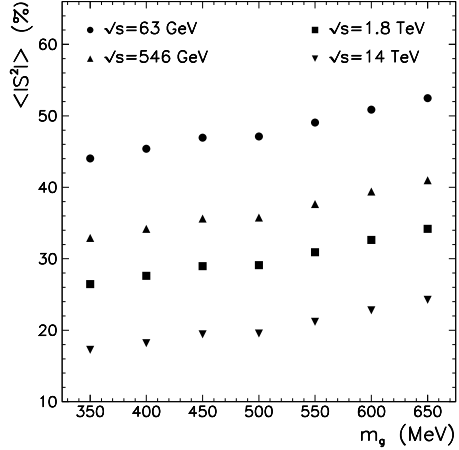}
\caption{The survival probability $\langle |S|^{2}\rangle$ (central values) as a function of the dynamical gluon mass $m_{g}$.}
\end{figure}

\section{Resummation of soft gluon radiation}

It is possible to apply the soft resummation
mechanism to the calculation of overlap functions in which the coupling $\alpha_{s}$ is constrained by the value
of the dynamical gluon mass. The main steps of this calculation can be summarized as follow: in QED the soft photon resummation in the energy-momentum
$K_{\mu}$ can be obtained order by order as \cite{godbole,godbole3}
\begin{eqnarray}
d^4P(K)=d^4K \int \frac{d^4x}{(2\pi)^4}\, e^{iK\cdot x-h(x,E)}, \hspace{0.5truecm} \textnormal{where}
\end{eqnarray}
\begin{eqnarray}
h(x,E) = \int_{0}^{E} d^3\bar{n}(k) [1-e^{-ik\cdot x}] = \int_{0}^{E} \frac{d^3k}{2k_{0}}\, \left| j_{\mu}(k,\{p_{i}\})\right|^{2} ;
\end{eqnarray}
here $d^4P(K)$ is the four-dimensional probability distribution for soft massless quanta emitted by a
semiclassical source, $E$ is the maximal energy allowed to single photon emission and $\{p_{i}\}$ is the momenta
of the emitting fields.

For strong interactions the resummed transverse momentum distribution is given by
\begin{eqnarray}
d^2P({\bf K}_{\perp})=d^2{\bf K}_{\perp} \int \frac{d^2{\bf b}}{(2\pi)^2}\, e^{-i{\bf K}_{\perp}\cdot {\bf b}-h(b,E)} ;
\end{eqnarray}
in QCD applications $h(b,E)$ can be written as \cite{qcd1,qcd2}
\begin{eqnarray}
h(b,E) = \frac{16}{3}\int^{E} \frac{dk_{t}}{k_{t}}\, \frac{\alpha_{s}(k_{t}^{2})}{\pi} \ln \left( \frac{2E}{k_{t}} \right)
\left[ 1-J_{0}(k_{t}b)  \right] ,
\end{eqnarray}
with the integral of the function $h(b,E)$, which describes the relative transverse momentum distribution
induced by soft gluon emission from a pair of initially collinear partons, downed to infrared momentum modes.

The attenuation of the rise of the total cross sections comes from soft gluon $k_{t}$-emission. These emissions
break collinearity between the colliding partons, hence changing the overlap function $W(b,s)$ (matter distribution), given by
\begin{eqnarray}
W(b,s) = N \int d^2{\bf K}_{\perp} \, e^{-i{\bf K}_{\perp}\cdot {\bf b}} \, \frac{d^2P({\bf K}_{\perp})}{d^2{\bf K}_{\perp}} =
W_{0}(s)\, e^{-h(b,q_{max})}. 
\end{eqnarray}
Note that the high-energy dependence of the total hadronic cross section (large $b$ value) is intrinsically related to the
small $k_{t}$ value of $\alpha_{s}$. Thus the behaviour of $\alpha_{s}$ in the infrared limit plays a central role.

In order to calculate $W(b,s)$ we have adopted the frozen-strong coupling $\bar{\alpha}_{s}$ obtained by Cornwall \cite{cornwall1,cornwall2},
\begin{eqnarray}
\bar{\alpha}_{s} (k_{t}^{2})= \frac{4\pi}{\beta_0 \ln\left[
(k_{t}^{2}+4M_g^2(k_{t}^{2}))/\Lambda^2 \right]}, \hspace{0.5truecm} \textnormal{where}
\end{eqnarray}
\begin{eqnarray}
M^2_g(k_{t}^{2}) = m_g^2 \left[ \frac{\ln \left( \frac{k_{t}^{2}+4{m_g}^2}{\Lambda^2} \right) }{\ln
\left( \frac{4m_g^2}{\Lambda^2} \right) } \right]^{- 12/11} .
\end{eqnarray}

We have compared our result with those obtained using the usual phenomenological prescription to the strong coupling
based on Richardson potential \cite{godbole,godbole3}:
\begin{eqnarray}
\alpha_{s}^{RCH} (k_{t}^{2})= \frac{4\pi}{\beta_0}\, \frac{p}{\ln\left[ 1 +
p \left( k_{t}^{2} / \Lambda^2 \right)^{p} \right]}.
\end{eqnarray}

The overlap functions $W^{DGM}(b,s)$ and $W^{RCH}(b,s)$, calculated using the expressions (15) and (17),
respectively, are displayed in the Fig. 3.
\begin{figure}
\includegraphics[width=0.70\columnwidth]{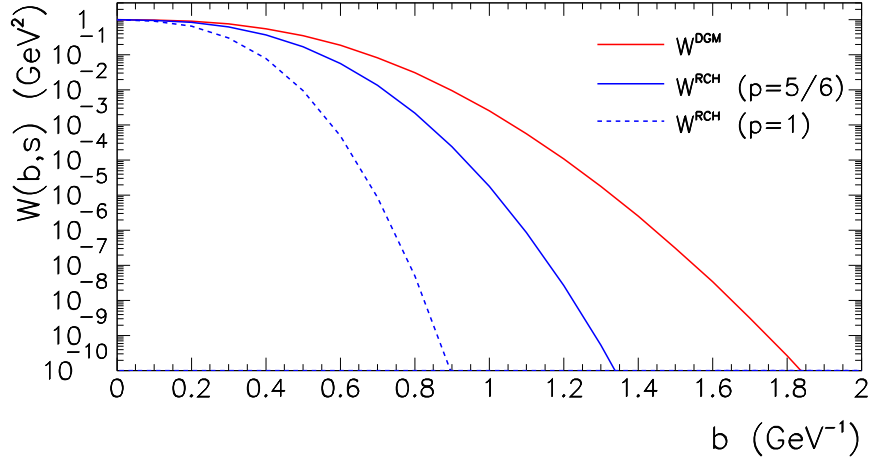}
\caption{The overlap functions calculated by means of $\bar{\alpha}_{s} $ and $\alpha_{s}^{RCH}$.}
\end{figure}

\section{Conclusion}

The frozen coupling $\bar{\alpha}_{s}$ avoids small $k_{t}$ divergences in partonic subprocesses. Thus it provides an useful
phenomenological tool to the study of high-energy strong interactions. The general picture indicates a smooth transition from nonperturbative
to perturbative behaviour of the QCD. Interestingly enough, by means of a triple-Regge analysis which explicitly accounts for
absorptive corrections, this smooth transition can also be inferred from recent determinations
of the bare triple-Pomeron vertex \cite{misha1,misha2}.

\end{document}